\documentstyle[aps,twocolumn]{revtex}

\begin{document}

\twocolumn[\hsize\textwidth\columnwidth\hsize\csname 
@twocolumnfalse\endcsname

\title{Deterministic walks in random media: evidence of generic
scale invariance}

\author{Gilson F. Lima$^{1,2}$,
Alexandre S. Martinez$^1$ and Osame Kinouchi$^1$\\
$^1$ 
Faculdade de Filosofia, Ci\^encia e Letras de Ribeir\~ao Preto, 
Universidade de S\~ao Paulo\\
Av. dos Bandeirantes 3900, CEP 14040-901, 
Ribeir\~ao Preto, SP, Brazil\\
$^2 $Escola T\'ecnica Federal de Mato Grosso,\\
R. Zulmira Canavarros 95, CEP 78005-390, Cuiab\'a, MT, Brazil.}

\maketitle

\begin{abstract}

Deterministic walks over a random set of points
in one and two dimensions ($d=1,2$) are considered.
Points (``cities'') are randomly scattered in ${\mathbf R}^d$
following a uniform distribution. 
A walker (a ``tourist''), at each time step, goes to the nearest neighbor city
that has not been visited in the past $\tau$ steps.
Each initial city leads to a trajectory composed of
a transient part and a final $p$-cycle attractor.
The distribution of transient times, $p$-cycles and 
number of cities per attractor are studied.
It is shown numerically that transient times (for $d=1,2$) follow a Poisson
law with a $\tau$ dependent decay
but the density of $p$-cycles follows a power law $D(p) \propto
p^{-\alpha(\tau)}$ for $d=2$. For large $\tau$, the
expoent tends to $\alpha\approx 5/2$. Some analytical results are
given for the $d=1$ case. Since the power law is robust
and does not depend on free parameters, this system presents 
``generic scale invariance''.
Applications to animal exploratory behavior and other local
minimization problems are suggested.

\end{abstract}

]

The study of random walks has been very fruitful in physics and mathematics, 
and the theory of such stochastic processes is a well developed subject. 
The study of deterministic walks is also an 
interesting subject but presents the analytical 
difficulties common to the area of 
non-linear dynamical systems and has been less investigated 
\cite{Ant}. Here we propose a simple and intriguing
problem, a deterministic walk over a random graph with $N$ nodes
that is also an example of a local (``on-line'') optimization dynamics. It 
may be called the ``local traveling salesman problem''
or perhaps the ``tourist problem'' for short.
The deterministic dynamics produces a division 
of the system phase space in several ${\mathcal O}(N)$
attractor basins which trap the walker (ergodicity is broken). 
The problem is reminiscent of walks in rugged landscapes but
the equivalent of ``local minima'' are cycles instead of point attractors.

The model is defined as follows: points are randomly distributed 
with a uniform density $\rho$ in ${\mathbf R}^d$,
where $d$ is the dimensionality of the space. 
These points may be thought as ``cities" or ``safe 
places" and they may be viewed as vertices of a random graph. 
At each time step, the ``tourist" follows the deterministic rule: 
{\bf Go to the nearest city (or place) that has not been 
visited in the past $\tau$ time steps.} Notice that the tourist wants 
to minimize only the distance to the next city (a local optimization
procedure), not the sum of all distances in the trajectory or some 
other global cost function. 

Our model can be of general interest for optimization 
theory with local constraints,
and studies of deterministic dynamical systems with quenched 
disorder. However, we would like to suggest some specific
motivations for considering this class
of problems. The local optimization procedure could be naturally related
to exploratory, foraging or migratory behaviors of animals.
For example, rodents present two competing drives: an exploratory
drive (``curiosity'') and a defensive behavior
called thigmotaxis. The later refers to rodent aversion to open spaces
and preference for places where its whiskers can touch vertical surfaces
or objects, which provide protection \cite{SMR}. 
It is arguable that, for biological agents (and biologically 
inspired robots) it could be sometimes more important to minimize the distance 
traveled in each movement between two safe places
instead of to optimize some global cost function.
In another scale, the model could describe migratory or
nomadic behaviors of humans, elephants, flamingos
and other animals with well developed spatial memory.
Cycles could be related to stable migratory routes on environments with 
localized resources, for example, oceanic islands, oasis and water holes. 
Local procedures are also usual in optimal foraging theory \cite{TSP}.
The need for local optimization emerges due to short range sensorial capacities. 
Long distances may also imply non-additive costs:
animals (and tourists) need safe places to stay during night, which 
puts a maximal distance that can be traveled at each time step.

Starting from a random city, the tourist performs a trajectory composed
of a transient part and a final $p$-cycle attractor.
In this letter we report the statistics for some relevant quantities similar 
to those measured in Kauffman networks \cite{BL}:
a) the probability $P_\tau(t)$ for obtaining a transient 
of size $t$, defined as the number of steps before the walker enters 
in some attractor (irrespective to the cycle period); for large $t$,
it is Poisson-like $P_\tau(t) \propto \exp(-t/\xi(\tau))$ with the decay time
$\xi(\tau)$ growing exponentially for $d=1$ and linearly for $d=2$; 
b) the total density of attractors ${\cal D}(\tau)$, which decays
exponentially for $d=1$ and as $\tau^{-1}$ for $d=2$; c) for $d=2$,
the density of $p$-cycles $D_\tau(p)$ which follows a power law $D_\tau(p)
\propto p^{-\alpha(\tau)}$ for $\tau>0$, with $\alpha\approx 5/2$ for 
large $\tau$ (generic scale invariance); and
d) the average number $\langle n_\tau(p)\rangle$ of cities present in a 
$p$-cycle. Also some analytical results are obtained for the 
$d=1$ case.

The discrete time step is simply a label: it does not measure
the actual physical time spent when the walker travels between the points.
This independence makes irrelevant the density $\rho$ of cities 
because only relative distances are important (which city is the nearest) and
not absolute distances, contrasting to standard random walks where
the mean length step defines an intrinsic length scale.
The only parameter is the memory window $\tau$.  
Self-avoidance is limited to this window and trajectories
can intersect outside this range. 
If $\tau=0$ (no memory) the tourist goes simply
to the nearest city until it finds two cities that
are reciprocally nearest neighbors, entering in a $2$-cycle. 
In this simple case, attractors may be identified with geometrical 
(cluster) properties. For $\tau = N - 1$ the trajectory is
totally self-avoiding and one 
has a kind of TSP nearest-neighbor algorithm \cite{TSP}. 
The interesting cases are the intermediate ones. For 
example, if $\tau = 1$, the last visited city cannot be revisited, and
only $p$-cycles with $p \geq 3$ can exist. For generic $\tau$, the
relation $p\geq \tau+2$ holds.

In the numerical experiments, $N$ points
are randomly scattered following a uniform 
distribution in the interval $[0,1]^d$. Each point has  
$d$ spatial coordinates $(x^1=x,x^2=y,\ldots,x^d)$.
The cities receive arbitrarily labeled
as $i=1,\ldots,N$ and one constructs the 
Euclidean distance matrix $\mathbf D$ (for example,
$D_{ij}=\sqrt{(x_i-x_j)^2+(y_i-y_j)^2}$ in $d=2$). 
The dynamics can be performed
over the entries of this matrix instead of on real space.
Starting from some city, one gets a transient trajectory until
the walker enters some periodic attractor and is trapped. 
The number of steps before the walker enters the cycle
defines the transient length $t$. The period $p$ and
the number of different cities $n$ that pertain to the attractor 
are also determined. The same city can be visited
more than once, thus $n\leq p$.

A finite size study showed that the behavior of the system is smooth as
a function of $1/N$, so we have used $N=3000$ as a reasonable number for
our simulations. Since each city 
is used as a starting point, a landscape with $N$ cities produces $N$ different 
transients. The statistics over $N_R$ realizations of sets of cities 
(``landscapes'') are collected. Unless stated otherwise, 
the results have been obtained by using $N_R = 100$.   

A natural question is if there is some critical $\tau$ that produces
a phase transition, for example the emergence of
an untraped (percolating) transient state.
The distribution of transient times does not suggest
this possibility because it is Poisson,
$P_\tau(t)\propto \exp(-t/\xi(\tau))$. This is shown for $d=1$ in
Fig.~1 and for $d=2$ in Fig.~2. The
exponential decay may be understood as follows. One finds numerically
that the total
number of attractors $N_A(\tau)$ is proportional to $N$, that is, 
the total density of attractors ${\cal D}(\tau)= N_A/N$ is constant 
and depends only on $\tau$. These attractors are scattered in phase space in
a random uniform manner. Like points scattered randomly in space, one expects 
that the distribution of distances between attractors follows a Poisson law
when these distances are larger than the attractor size $p$.
By supposing that transient times are proportional to
these distances, a Poisson law follows also for the transient times. 

For $d=1$, the characteristic times grow as
$\xi(\tau) \propto \exp(\gamma\tau)$
(inset Fig.~1) and the total density decays as
${\cal D}(\tau) \propto \exp(-\gamma^\prime \tau)$ (Fig.~3a).
For $d=2$, one observes the linear dependence 
$\xi((\tau) \propto \tau^\delta, \:\delta=1.0$ (inset Fig.~2); 
the total density decays as a power law ${\cal D}(\tau) \propto 
\tau^{-\delta^\prime}$ (Fig.~3b). The average transient is proportional 
to the average distance between attractors,
(which are inversely proportional to the attractor density). This
means that ${\cal D}(\tau) \cdot \xi(\tau)$
should be constant, that is, $\gamma=\gamma^\prime$ and
$\delta=\delta^\prime$. This is indeed the case (see Fig.~3c),
but for $d=1$, although the exponential terms cancel, 
a linear residue remains. A better expression for the $1D$
decay time is $\xi(\tau) = c \tau \exp(\gamma \tau)$. We conjecture
that the linear prefactor arises from the transient time
spent in the cities of the attractor before it stabilizes
(this time is larger for $d=1$ systems). 
 
A property of natural interest is the
density $D_\tau(p)$ of $p$-cycles, estimated as the
number of {\em different\/} $p$-cycles divided by $N$, 
in the limit of very large systems. Evaluating this quantity
requires careful enumeration because, 
when starting from all the possible initial states, one must not count 
the same attractor twice. Notice that ${\cal D}(\tau)= \sum_p
D_\tau(p)$. For $d=1$, $D_\tau(p)$ is certainly non Poisson, 
although the evidence for a power law is weak (Fig.~4a). 
For $d=2$, one observes  
clear power laws $D_\tau(p) \propto p^{-\alpha(\tau)}$ (Fig.~4b).
The exponent $\alpha(\tau)$ stabilizes around $\alpha=5/2$ for large 
$\tau$ (inset Fig.~4b). Since there is no fine tuning of any explicit 
parameter in our system (such as $\tau$), the scale invariance is ``generic''
\cite{Grinstein}, that is, intrinsic to the
problem. This is the most surprising result of our study.
We conjecture that this scale invariance is related to 
two factors: a) a uniform distribution of points has a single length scale, 
$\lambda=\rho^{1/d}$ but this length is irrelevant to 
the dynamics since only relative distances are considered when making 
a move; and b) the window $\tau$ defines a minimal length $p_{min}= \tau+2$ 
but not a maximal one. A clear explanation of this power law
is still lacking.

We stress that this problem is not related only
to geometrical properties since the cycles are appear only due to the 
introduced dynamics. Naively, one could think 
that a $p$-cycle is a geometrical object, for example a cluster where
the distances between the points are smaller than any distance outside
the cluster. This indeed is a sufficient but not necessary condition to obtain a 
$p$-attractor. For example, for $d=2$ (Fig.~5a), 
a walker with memory $\tau =1$ starts from city $A$ and finds the 
$4$-cycle $ABCD$. Although city $E$ 
is close to the cluster (since $BE<AB$), it is never visited because 
$BC < BE$ and $CD<CE$. However, if the tourist starts from city $C$, 
one gets a $3$-cycle that includes city $E$!
This degeneracy and superposition of attractors can be understood 
observing that Fig.~5 shows trajectories in configuration space, 
not in phase space. In phase space, points corresponds to 
$\tau+1$-uples $({\bf X}_t,\ldots, {\bf X}_{t-\tau})$
where ${\bf X}_t$ is the position $(x,y)$ of the 
tourist at time $t$ and trajectories never intersect. Only for 
$\tau=0$ the configuration space is equivalent to the phase space. 

Finally, we present the average number of cities $\langle n_\tau \rangle(p)$ 
pertaining to cycles of period $p$ (Fig.~6). For $d=1$ 
there is almost no dispersion in the number of cities per attractor. 
A $p$-cycle has $n(p)$ cities. We also found that, 
for $n>2(\tau+2)$, the following relation holds for even cycles:
\begin{equation}
n_\tau(p)= p/2 +\tau +1 \:. \label{pn}
\end{equation}

To see how this relation emerges,
notice that for each $\tau$ there
is a minimum cycle of period $p_\tau=\tau+2$, which we call
a {\em base block\/} (Fig.~5b). A base block is composed of
$n_\tau=\tau+2$ cities. 
The next cycles follow specific constructions (Fig.~5c). 
But when $n$ is large, geometrical constraints impose that 
the most common $p$-cycles are made of two base blocks
(one in each attractor extremity) joined by $n_I$ {\em intermediate cities\/},
see Fig.~5d. An attractor with $n$ cities thus have 
$n_I=n-2n_\tau$ intermediate points. 
These intermediate cities contribute to the total period with 
$p_I=2n_I+2$ steps (since for $n_I=0$, the joining of the base blocks 
contributes with two steps). Thus, the total period is $p= 2\times 
p_\tau + p_I = 2(n-\tau-1)$, which leads to Eq.~(\ref{pn}).
This relation holds for cycles with $n\geq 2 n_\tau = 2(\tau+2)$, 
because only these cycles can incorporate
two independent base blocks. 
For $\tau=1$, this is the unique conceivable manner of constructing 
cycles, meaning that odd cycles are prohibited (and also $p=6$ cycles, see Fig.~4a). For $\tau >1$, it is possible to construct odd 
cycles by using internal loops 
(an example with $\tau=2$ is given in Fig.~5e). 

In $d=2$, the attractors are
polygons with different forms and shapes so that this strict relation 
between periods and cities does not hold, although $\langle n \rangle$ also scales
linearly with $p$ (not shown). 
For $\tau=1$, one finds that
odd cycles are less probable than even cycles (Fig.~4b), which is reminiscent
of the $d=1$ behavior. Indeed, this occurs because elongated
odd attractors in two-dimensional space are prohibited 
by the same geometrical constraints present in the 
one-dimensional case. 
     
Another analytical result for the $d=1$ case can be obtained. 
Consider points $x_i$ randomly scattered
along the real line, defining segments of size $s_i=x_i-x_{i-1}$.
Without loss of generality, we assume that $\rho=1$, which means 
that
 $\langle s \rangle = 1$. It is easy to see that the distribution 
of interval sizes $P(s)$ follows a Poisson distribution 
$P(s)= \Theta(s)\exp(-s) $, where $\Theta(s)$
is the Heaviside step function. For $\tau=0$, there 
exist only $2$-cycles attractors, which correspond to pairs of
reciprocal nearest neighbors. The probability $P_2$ for 
this configuration is equal to the probability that $s_{i-1} > s_i$ 
{\bf and} $s_{i+1} > s_i$. Since $s_{i-1}$ and $s_{i+1}$ 
are drawn independently, one gets:
$P_2 = \int_{-\infty}^\infty ds_i \: P(s_i)\:
P(s_{i-1}>s_i | s_i) \:P(s_{i+1}>s_i |s_i)  
=  \int_0^\infty ds_i \:\mbox{e}^{-s_i} \left(\int_{s_i}^\infty ds\: 
\mbox{e}^s \right)^2  = 1/3 $,
that is, on average, one third of the sequences of four points leads to
reciprocal nearest neighbors and so to $2$-cycles. 
Since the number of sequences of four points is, in the large $N$ 
limit, equal to the number of points,
one obtains $D_0(2)= 1/3$. This has been fully confirmed by our 
numerical simulations (see caption Fig.~4). 

The model may be generalized by introducing
a stochastic component (a ``temperature'' $T=1/\beta$). 
For example, the probability for the tourist to travel 
from its present city $j$ to some
city $i$ may be a function of the distance, say, 
$P(j\rightarrow i) \propto \exp(-\beta D_{ij}/\lambda)$, 
where $\lambda=\rho^{1/d}$ normalizes the distances. 
In this case we expect, for $T=\beta^{-1}\ll 1$, a 
punctuated-equilibrium behavior with
sporadic transitions between attractor basins. It is an open question
to determine if there is a critical temperature $T_c$ where full 
ergodicity is recovered. 

{\bf Acknowledgments:} O. Kinouchi thanks N. Caticha and the
participants of the Statistical Physics of Neural Networks seminar
at the Max Plank Institute for the Physics of Complex Systems,
Dresden (march 1999) for useful discussions when this problem was being
formulated. R. Vicente, N. Alves and U. P. C. Neves made useful
suggestions for the final version of the manuscript. We also 
thank P. Stadler for discussing the problem and for the
suggestion of the name ``tourist problem'', made during his visit in Brazil.
O. Kinouchi acknowledges support from FAPESP.

%
%
%
%

%
%
%
%
%
\newpage

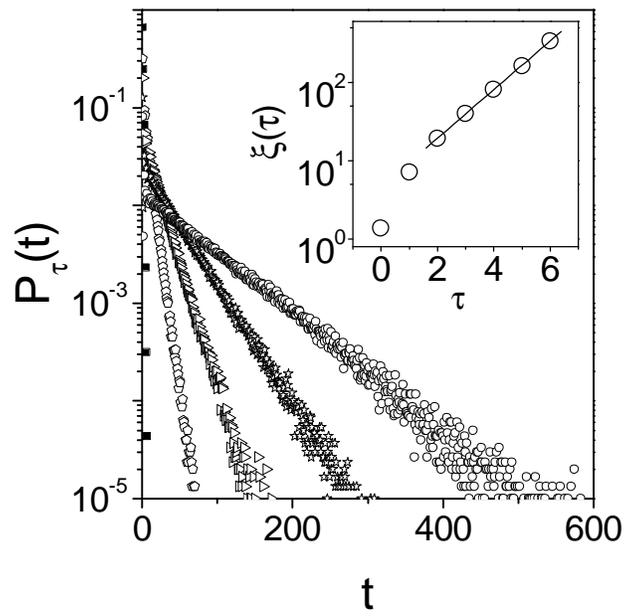
\begin{figure}
\caption{
Distribution of transient times $P_\tau(t)$ for $d=1$. 
From right to left: $\tau= 5, 4, 3, 2$ and $0$ (squares near the $y$ axis). 
Inset: Decay time $\xi(\tau)$. 
}
\end{figure}

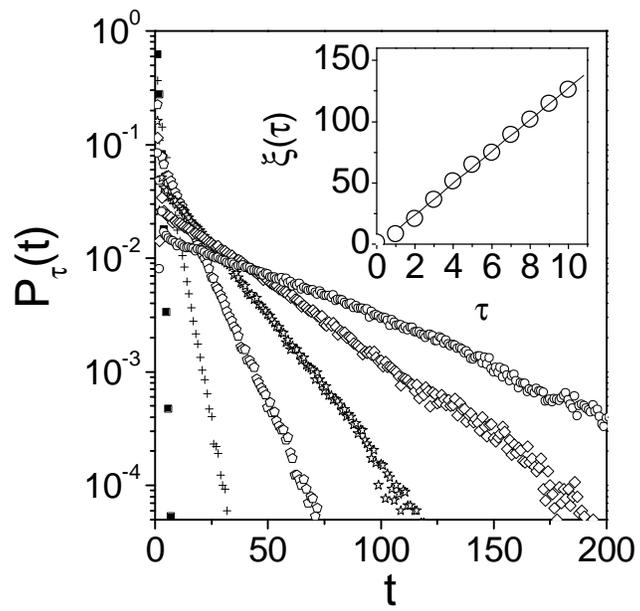
\begin{figure}
\caption{
Distribution $P_\tau(t)$ of transient times 
for $d=2$. From right to left:
$\tau= 10, 5, 3, 2, 1$ and $0$. Inset: Decay time $\xi(\tau)$. 
}
\end{figure}

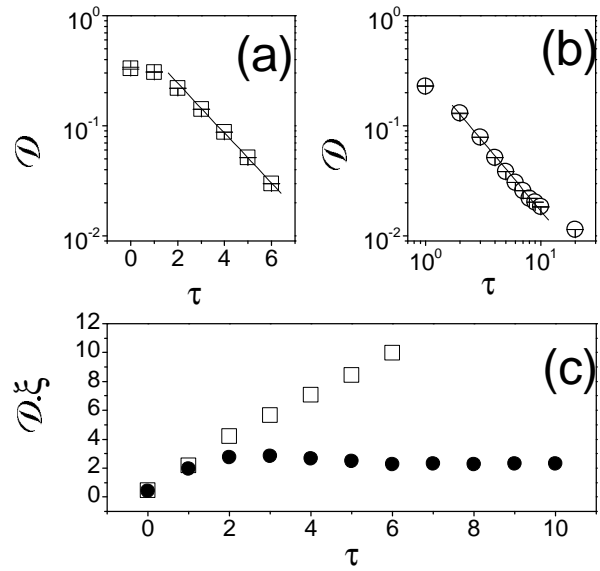
\begin{figure}
\caption{
Total density of attractors ${\cal D}(\tau)$: 
a) $d=1$; b) $d=2$; 
c) ${\cal D}(\tau)\cdot\xi(\tau)$ for $d=1$ (squares) and $d=2$ (circles),
error bars smaler than symbol size.
}
\end{figure}

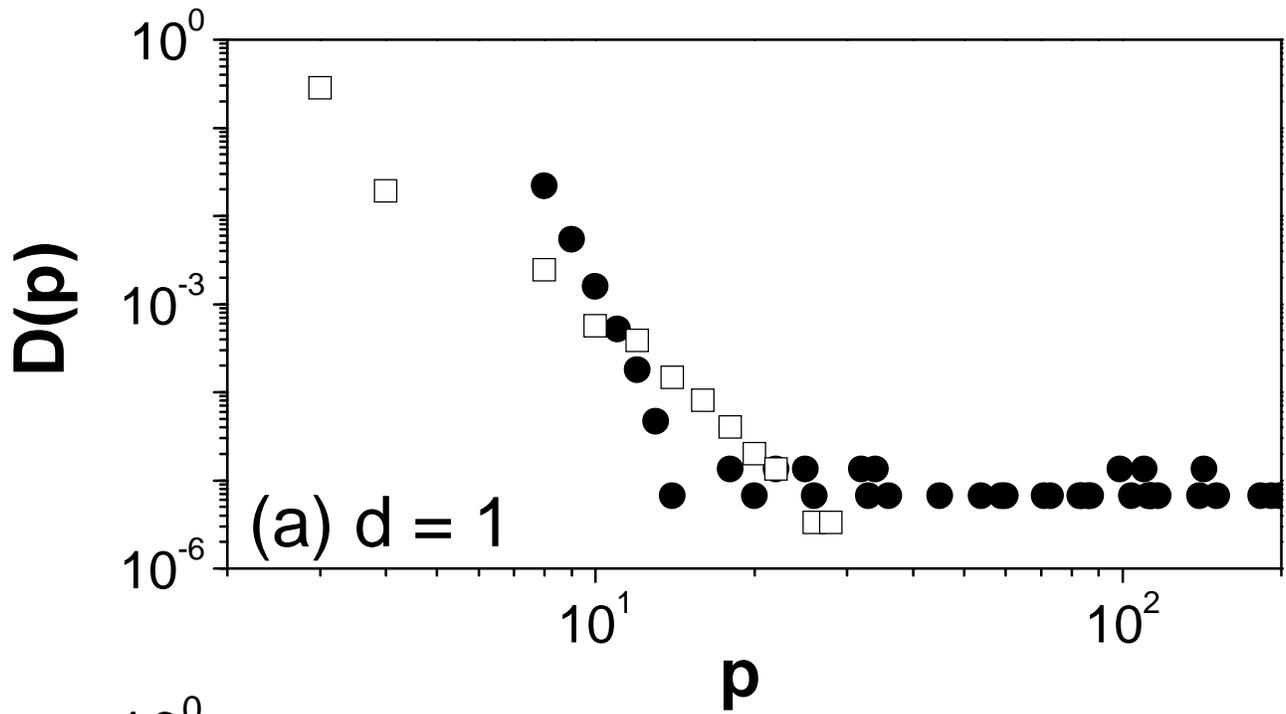
\begin{figure}
\caption{
Examples of attractor densities ${\cal D}_\tau(p)$: a) $d=1$, 
$\tau =1$ (squares) and $\tau = 6$ (filled circles);
b) $d=2$, $\tau=1$ (squares) and $10$ (filled circles), $N=3000$ and
$N_R=700$. Inset: exponent $\alpha(\tau)$. 
For $\tau=0$ only $2$-cycles exist: $D_0(2)= 0.333 \pm 0.001$, 
for $d=1$ and $D_0(2)= 0.31 \pm 0.01$ for $d=2$. 
}
\end{figure}

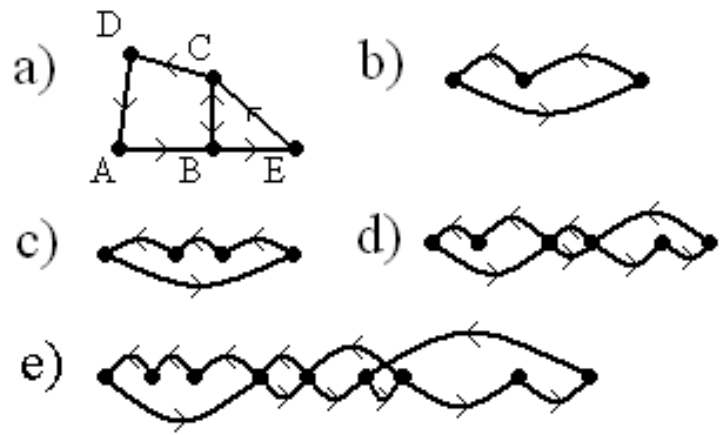
\begin{figure}
\caption{
 a) Example of superposition of attractors for $d=2$ and
$\tau =1$: starting from $A$ one obtains the $4$-cycle $ABCD$, 
but starting on $C$ one gets the $3$-cycle $CBE$; b) the $3$-cycle
base block for $\tau=1$; c) the $4$-cycle for $\tau=1$; 
d) next permissible cycle for $\tau=1$:
a $8$-cycle made of two base blocks; e) example of 
odd ($p=13$) cycle for $\tau=2$ which is possible because of
an internal loop.
}
\end{figure}

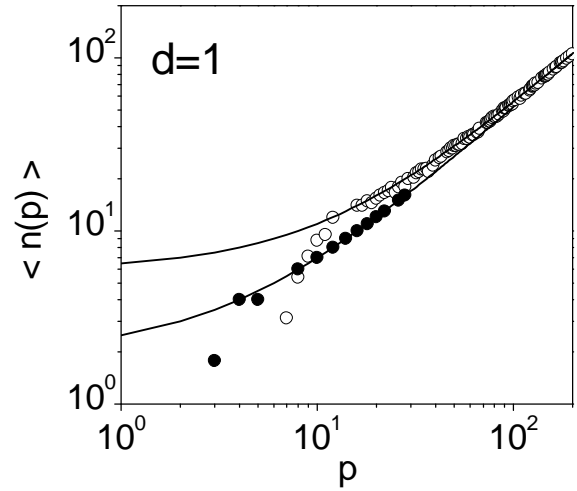
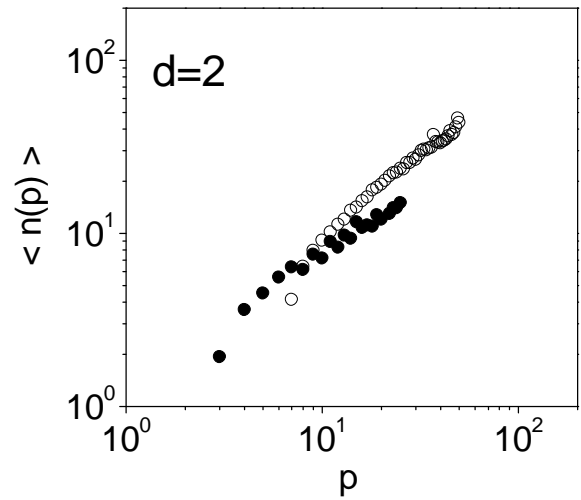
\begin{figure}
\caption{
Number of cities per attractor $\langle n(p) \rangle$. 
a) $d=1$, $\tau=1$ (filled circles) and $\tau=5$ (circles),
theoretical curves (solid) $n(p) = p/2 +\tau +1$. 
}
\end{figure}

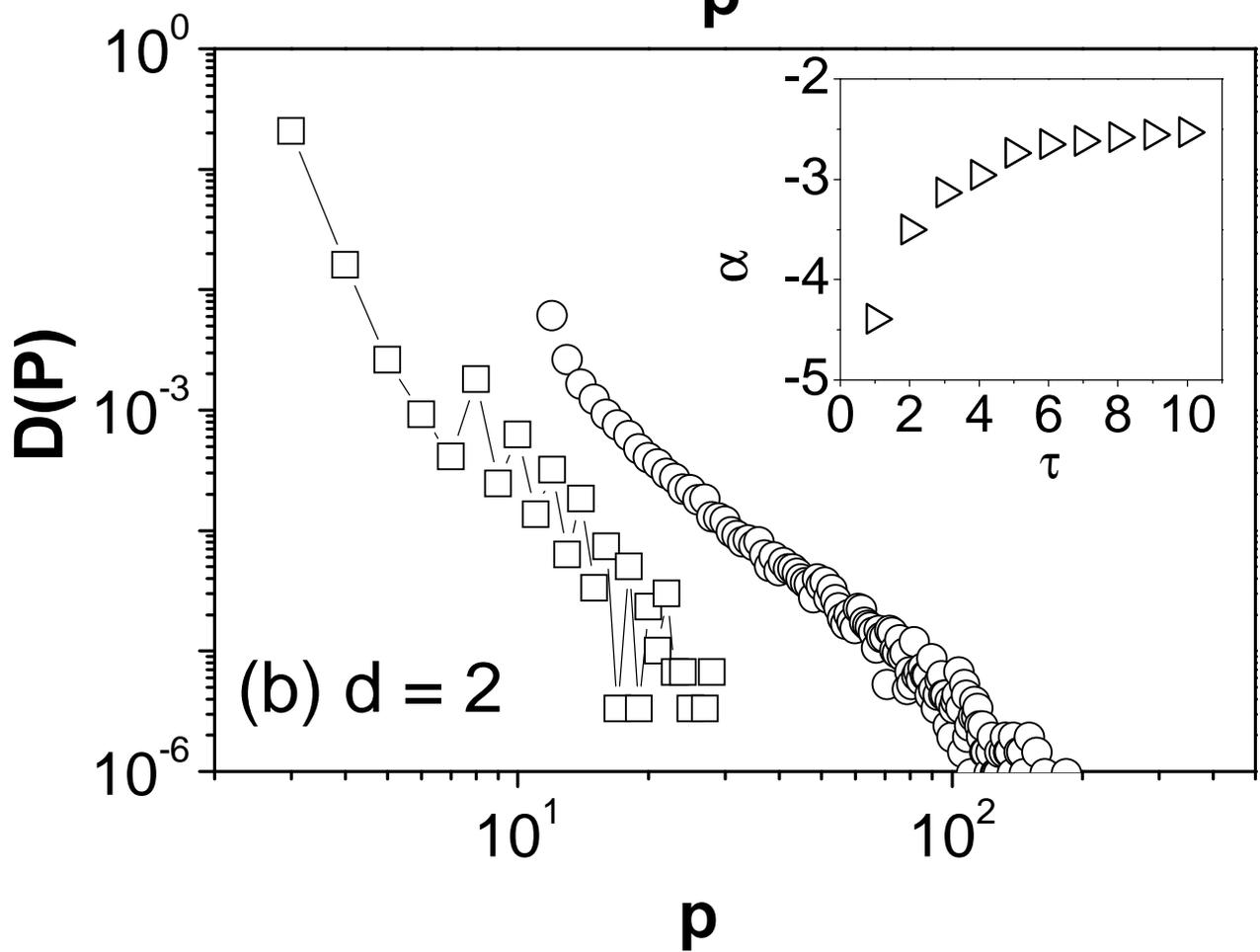
\end{document}